\documentclass[submission,copyright,creativecommons]{eptcs}
\providecommand{\event}{ACT 2019} 
 \usepackage{breakurl}             

\usepackage[utf8]{inputenc}
\usepackage[english]{babel}
\usepackage[T1]{fontenc}
\usepackage{amsmath,amssymb,amsthm}
\usepackage{hyperref}
\usepackage[numbers,sort&compress]{natbib}
\usepackage{lipsum}
\usepackage[all,2cell,cmtip]{xy}
\UseTwocells
\usepackage[svgnames]{xcolor}
\usepackage{caption}
\usepackage{subcaption}

\usepackage{tikz}
\usetikzlibrary{decorations.markings}
\usetikzlibrary{shapes.geometric}
\usetikzlibrary{cd}

\tikzstyle{arrowhead}=[-latex,draw=black,line width=2.000]
\tikzstyle{wbox}=[rectangle,fill=white,draw=black, inner sep=1.5pt]
\tikzstyle{species}=[circle,fill=white,draw=black,scale=.75]
\definecolor{lblue}{rgb}{0,250,255}
\tikzstyle{transition}=[rectangle,fill=lblue,draw=black,scale=.75]
\tikzstyle{arrow}=[-,postaction={decorate},decoration={markings,mark=at position .7 with {\arrow{>}}},line width=1.00]
\tikzstyle{simple}=[-,draw=black,line width=1.000]
\tikzstyle{ar}=[->,shorten >=1.2pt]

\tikzstyle{bus}=[circle,fill=black,draw=black,scale=0.4]
\tikzstyle{equation}=[circle,draw=black, fill=none,scale=0.4]
\tikzstyle{none}=[circle, inner sep=0]
\tikzstyle{junction}=[circle,fill=black,scale=0.18]

\pgfdeclarelayer{edgelayer}
\pgfdeclarelayer{nodelayer}
\pgfsetlayers{edgelayer,nodelayer,main}

\newcommand{\R}{\mathbb{R}}
\newcommand{\CC}{\mathbb{C}}

\newcommand{\D}{\mathcal{D}}
\newcommand{\DiGraph}{\mathsf{DiGraph}}
\newcommand{\Markov}{\mathsf{Markov}}

\newcommand{\Set}{\mathsf{Set}}

\newcommand{\DER}{\mathsf{DER}}
\newcommand{\DERMarkov}{\mathsf{ DER_{Mark} }}
\newcommand{\DERBase}{\mathsf{ DER_{Base} }}
\newcommand{\DERCost}{\mathsf{DER_{Cost} }}
\newcommand{\Weather}{\mathsf{Weather}}
\newcommand{\Inst}{\mathsf{-Inst}}
\newcommand{\eval}{\mathsf{eval}}
\newcommand{\coeval}{\mathsf{coeval}}
\DeclareMathOperator{\id}{id}

\newcommand{\maps}{\colon}

\theoremstyle{plain}

\newtheorem{thm}{Theorem}

\newtheorem{defn}[thm]{Definition}

\theoremstyle{remark}

\title{Compositional Models for Power Systems}
\date{}

\author{John S.\ Nolan 
\institute{University of Maryland}
\and Blake S.\ Pollard \quad Spencer Breiner  \\  Dhananjay Anand 
\institute{National Institute of Standards and Technology}
\and
Eswaran Subrahmanian
\institute{Carnegie Mellon University}
}

\def\authorrunning{Nolan, Pollard, Breiner, Anand, Subrahmanian }
\def\titlerunning{Compositional Models for Power Systems}

\begin{document}
\maketitle

\begin{abstract}
The problem of integrating multiple overlapping models and data is pervasive in engineering, though often implicit. We consider this issue of model management in the context of the electrical power grid as it transitions towards a modern `Smart Grid.' We present a methodology for specifying, managing, and reasoning within multiple models of distributed energy resources (DERs), entities which produce, consume, or store power, using categorical databases and symmetric monoidal categories. Considering the problem of distributing power on the grid in the presence of DERs, we show how to connect a generic problem specification with implementation-specific numerical solvers using the paradigm of categorical databases.
\end{abstract}

\section{Introduction}
The modeling of complex systems, engineered or natural, entails certain generic challenges: the existence and interaction of multiple models, multiple algorithms, and multiple implementations. This paper presents a methodology rooted in category theory to manage this complexity, concretized via a model-driven engineering approach to designing a modern electrical grid, dubbed the `Smart Grid.'

The existing grid architecture is characterized by dedicated large-scale, centralized generation and distributed, downstream consumption. Moving towards an architecture with increased distributed generation will have a profound impact on how the grid is managed: end users will no longer be dedicated consumers, but will shift between consuming and producing power. One key to enabling this transition is the management and modeling of distributed energy resources (DERs), generic devices that can consume, produce, or store power. 

The notion of DER is meant to provide an abstraction or characterization summarizing the essential properties of a wide array of different energy resources, e.g.\ photovoltaic systems, batteries, conventional loads, and so on. The issue is that no uniform abstraction exists. Different stakeholders utilize different abstractions for different purposes. In addition, these meta-models must evolve as new technologies emerge.

Coupled with control mechanisms, DERs provide a number of ancillary services to consumers and grid operators: voltage control, reducing peak loads, demand response, etc.\ \cite{peak, rahimi2010demand}. Aggregations of heterogeneous DERs provide the abstraction through which such collections participate in the overall power system and energy markets \cite{chalkiadakis2011cooperatives}. 

There is a large body of work concerned with the use of model transformations in the context of model management and model-driven engineering, e.g.\ \cite{mens2006taxonomy, Stevens:2007:LBM:1462618.1462630}. A number of approaches utilize category theory, recognizing the natural mathematical framework it provides for reasoning about models, their semantics, and structure-preserving transformations among models \cite{diskin2014category, trollman2015}.

We tackle the problem of specifying, relating, and transforming models using the functorial data model advocated in \cite{johnson2002entity,spivak2012functorial, schultz2016databases, sw2017integration, functorial} as well as its computational implementation in the CQL tool. In the functorial data model, database schemas are interpreted as finite presentations of categories. Instances of a database schema correspond to $\Set$-valued functors out of the associated category. Some subtleties arise when working with computational data such as strings and integers, though we will not concern ourselves with these difficulties; see \cite{schultz2016databases} for a thorough treatment. 


{\bf Structure of this paper:} In Section \ref{sec:DER}, we consider the problem of model specification and transformation for family of DER schemas and the functors among them. 
In Section \ref{sec:DERstructure}, we show that one variant of a model in this family yields the objects of a symmetric monoidal category with DER aggregation as monoidal product. In Section \ref{sec:PF}, we consider the problem of distributing power in a grid where certain nodes correspond to aggregate collections of DERs, describing a procedure for translating among numerical solvers using database schemas, functors, and queries.

\section{Categorical databases for model management} \label{sec:DER}

The family of models we present in this section share a common ancestor, the directed multigraph, henceforth graph, consisting of two entities, States and Transitions, together with two arrows, Source and Target, assigning source and target states to transitions:
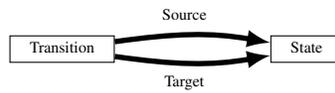
\[
\begin{tikzpicture}[scale=.65, font=\scriptsize]
	\begin{pgfonlayer}{nodelayer}
		\node [style=wbox] (1) at (-5, 0) {\tiny \begin{tabular}{c} Transition  \end{tabular}};
		\node [style=wbox] (3) at (0, 0) {\tiny \begin{tabular}{c} State \end{tabular}};
		\node (00) at (-2.5, 0.7) {\tiny Source};
		\node (01) at (-2.5, -0.7) {\tiny Target};

	\end{pgfonlayer}
	
	\begin{pgfonlayer}{edgelayer}
		\draw [style=arrowhead, out=8, in=170] (1) to (3);
		\draw [style=arrowhead, out=-8, in =190] (1) to (3);
	\end{pgfonlayer}
 \end{tikzpicture}
\]
Including identities and composites, this schema forms a category $\DiGraph$. Functors from this category to $\Set$ form a category of instances. 

\subsection{A basic DER model} \label{sec:modeltranslation}
 In our base model, DERs are viewed as graphs with operational states as nodes and transitions among those. Each state is assigned a feasible operating region (i.e.\ power demands / generation). In AC circuits, power is a complex-valued quantity $P+iQ$, where the real part $P$ is referred to as real or active power and the imaginary part $Q$ as reactive power. For now we restrict our attention to the case where operating regions are single points. 
 
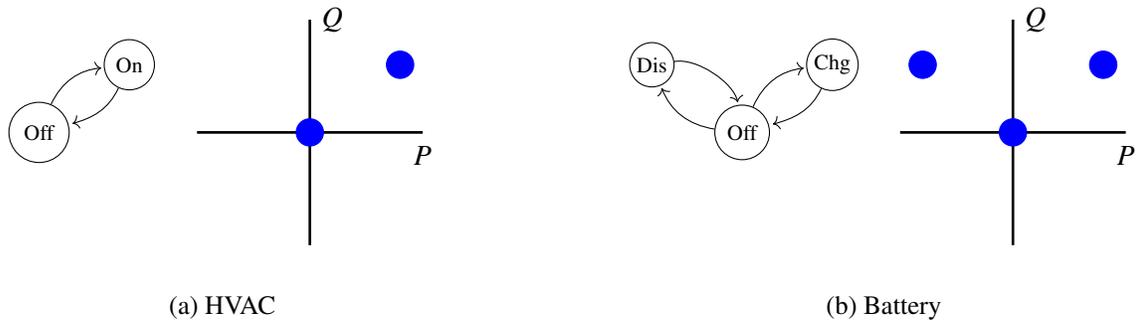
\begin{figure}[!h]

\begin{subfigure}[b]{0.45 \linewidth}
\[ \begin{tikzpicture}[scale=0.6]

\begin{pgfonlayer}{nodelayer}
		\node [style=species] (0) at (2, 1.5) {On};

		\node [style=species, inner sep = 5pt] (6) at (0, -0) {Off};

	\end{pgfonlayer}
	\begin{pgfonlayer}{edgelayer}

		\draw [->, shorten >=1pt, bend left, looseness=1.00] (6) to (0);

		\draw [->, shorten >=1.5pt, out=245, in=15] (0) to (6);
	\end{pgfonlayer}

\begin{scope}[shift= {(6,0)} ]

\begin{pgfonlayer}{nodelayer}

		\node [style=none] (1) at (0.5, 2.5) {$Q$};
		\node [style=none] (2) at (2.5, -0.5) {$P$};
		\node [style=none] (3) at (2.5, -0) {};
		\node [style=none] (4) at (0, 2.5) {};

		\node [circle, draw=blue, fill=blue] (5) at  (0,0) {};
		\node [circle, draw=blue, fill=blue] (5) at  (2 , 1.5) {};

		\node [style=none] (7) at (-2.5, -0) {};
		\node [style=none] (8) at (0, -2.5) {};
	\end{pgfonlayer}
	\begin{pgfonlayer}{edgelayer}
		\draw [style=simple] (4.center) to (8.center);
		\draw [style=simple] (7.center) to (3.center);
	\end{pgfonlayer}
\end{scope}

\end{tikzpicture} \]
\caption{HVAC}
\end{subfigure} \hfill
\begin{subfigure}[b]{0.45 \linewidth}
\[ 
\begin{tikzpicture}[scale=0.6 ]

\begin{pgfonlayer}{nodelayer}
		\node [style=species, inner sep= 2pt] (0) at (2, 1.5) {Chg};

		\node [style=species, inner sep = 2pt] (5) at (-2, 1.5) {Dis};
		\node [style=species, inner sep = 4pt] (6) at (0, -0) {Off};

	\end{pgfonlayer}
	\begin{pgfonlayer}{edgelayer}

		\draw [->, shorten >=1pt, bend left, looseness=1.00] (6) to (0);
		\draw [->, shorten >=1pt, bend left, looseness=1.00] (6) to (5);
		\draw [->, shorten >=1pt, bend left=45, looseness=0.75] (5) to (6);
		\draw [->, shorten >=1.5pt, out=245, in=15] (0) to (6);
	\end{pgfonlayer}

\begin{scope}[shift={(6,0)}]
\begin{pgfonlayer}{nodelayer}

		\node [style=none] (1) at (0.5, 2.5) {$Q$};
		\node [style=none] (2) at (2.5, -0.5) {$P$};
		\node [style=none] (3) at (2.5, -0) {};
		\node [style=none] (4) at (0, 2.5) {};

		\node [circle, draw=blue, fill=blue] (5) at  (0,0) {};
		\node [circle, draw=blue, fill=blue] (5) at  (2, 1.5) {};
		\node [circle, draw=blue, fill=blue] (5) at  (-2, 1.5) {};

		\node [style=none] (7) at (-2.5, -0) {};
		\node [style=none] (8) at (0, -2.5) {};
	\end{pgfonlayer}
	\begin{pgfonlayer}{edgelayer}
		\draw [style=simple] (4.center) to (8.center);
		\draw [style=simple] (7.center) to (3.center);
	\end{pgfonlayer}
\end{scope}
\end{tikzpicture} \]
\caption{Battery}
\end{subfigure}

\caption{Two types of DERs and their associated demand profiles. With the chosen convention, states with positive real power $P$ consume power, while states with negative $P$ generate power. }
\label{fig:DER}
\end{figure}

 Our base DER model is described by the schema $\DERBase$
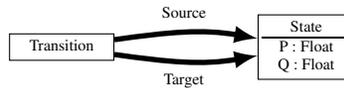
\[
\begin{tikzpicture}[scale=.65, font=\scriptsize]
	\begin{pgfonlayer}{nodelayer}
		\node [style=wbox] (1) at (-5, 0) {\tiny \begin{tabular}{c} Transition  \end{tabular}};
		\node [style=wbox] (3) at (0, 0) {\tiny \begin{tabular}{c} State \\ \hline P : Float \\ Q : Float \end{tabular}};
		\node (00) at (-2.5, 0.7) {\tiny Source};
		\node (01) at (-2.5, -0.7) {\tiny Target};

	\end{pgfonlayer}
	
	\begin{pgfonlayer}{edgelayer}
		\draw [style=arrowhead, out=8, in=170] (1) to (3);
		\draw [style=arrowhead, out=-8, in =190] (1) to (3);
	\end{pgfonlayer}
 \end{tikzpicture}
\]
consisting of a graph together with two attributes for each state, $P,Q \maps \text{State} \to \text{Float}$. Two instances of $\DERBase$ are depicted in Figure \ref{fig:DER}, showing a typical load (an HVAC, i.e.\ heating/cooling system) and a battery. 
 
\subsection{Model Translation via Functors}
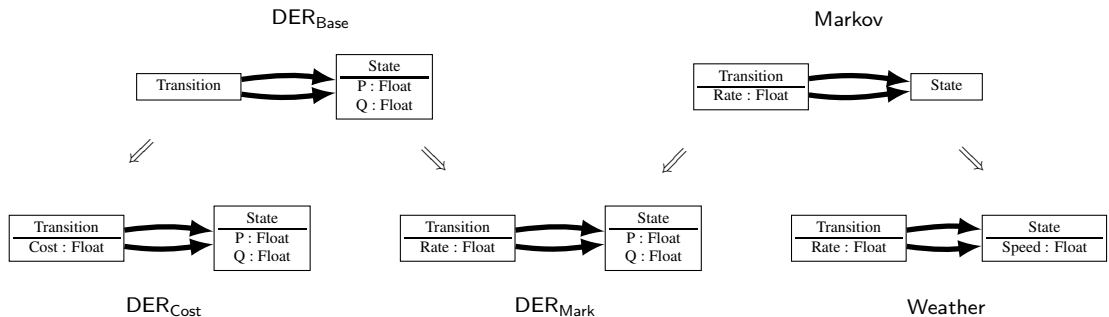
\begin{figure}[h]
\[
\begin{tikzpicture}[scale=.65, font=\scriptsize]

\begin{scope}[shift={(-1,0)}]
	\begin{pgfonlayer}{nodelayer}
	    \node (0) at (-2,1.4) {$\DERBase$};
		\node [style=wbox] (1) at (-4.5, 0) {\tiny \begin{tabular}{c} Transition  \end{tabular}};
		\node [style=wbox] (3) at (-0.5, 0) {\tiny \begin{tabular}{c} State \\ \hline P : Float \\ Q : Float \end{tabular}};
		\node [rotate=225] (00) at (-5.5, -1.3) {$\Longrightarrow$};
		\node [rotate=-45] (01) at (0.5, -1.5) {$\Longrightarrow$};

	\end{pgfonlayer}
	
	\begin{pgfonlayer}{edgelayer}
		\draw [style=arrowhead, out=8, in=172] (1) to (3);
		\draw [style=arrowhead, out=-8, in =187] (1) to (3);
	\end{pgfonlayer}
\end{scope}

\begin{scope}[shift={(-4,-3.1)}]
	\begin{pgfonlayer}{nodelayer}
		\node (0) at (-2,-1.4) {$\DERCost$};
		\node [style=wbox] (1) at (-4, 0) {\tiny \begin{tabular}{c} Transition \\ \hline Cost : Float  \end{tabular}};
		\node [style=wbox] (3) at (0, 0) {\tiny \begin{tabular}{c} State \\ \hline P : Float \\ Q : Float \end{tabular}};
	\end{pgfonlayer}
	
	\begin{pgfonlayer}{edgelayer}
		\draw [style=arrowhead, out=8, in=172] (1) to (3);
		\draw [style=arrowhead, out=-8, in =187] (1) to (3);
	\end{pgfonlayer}
\end{scope}

\begin{scope}[shift={(4,-3.1)}]
	\begin{pgfonlayer}{nodelayer}
		\node (0) at (-2,-1.4) {$\DERMarkov$};
		\node [style=wbox] (1) at (-4, 0) {\tiny \begin{tabular}{c} Transition \\ \hline Rate : Float  \end{tabular}};
		\node [style=wbox] (3) at (0, 0) {\tiny \begin{tabular}{c} State \\ \hline P : Float \\ Q : Float \end{tabular}};
	\end{pgfonlayer}
	
	\begin{pgfonlayer}{edgelayer}
		\draw [style=arrowhead, out=8, in=172] (1) to (3);
		\draw [style=arrowhead, out=-8, in =187] (1) to (3);
	\end{pgfonlayer}
\end{scope}

\begin{scope}[shift={(10,0)}]
\begin{pgfonlayer}{nodelayer}
	    \node (0) at (-2,1.4) {$\Markov$};
		\node [style=wbox] (1) at (-4, 0) {\tiny \begin{tabular}{c} Transition \\ \hline Rate : Float  \end{tabular}};
		\node [style=wbox] (3) at (0, 0) {\tiny \begin{tabular}{c} State \end{tabular}};
		\node [rotate=225] (00) at (-5.6, -1.5) {$\Longrightarrow$};
		\node [rotate=-45] (01) at (0.5, -1.5) {$\Longrightarrow$};

	\end{pgfonlayer}
	
	\begin{pgfonlayer}{edgelayer}
		\draw [style=arrowhead, out=8, in=172] (1) to (3);
		\draw [style=arrowhead, out=-8, in =187] (1) to (3);
	\end{pgfonlayer}
\end{scope}

\begin{scope}[shift={(12,-3.1)}]
	\begin{pgfonlayer}{nodelayer}
		\node (0) at (-2,-1.4) {$\Weather$};
		\node [style=wbox] (1) at (-4, 0) {\tiny \begin{tabular}{c} Transition \\ \hline Rate : Float  \end{tabular}};
		\node [style=wbox] (3) at (0, 0) {\tiny \begin{tabular}{c} State \\ \hline Speed : Float \end{tabular}};
	\end{pgfonlayer}
	
	\begin{pgfonlayer}{edgelayer}
		\draw [style=arrowhead, out=8, in=172] (1) to (3);
		\draw [style=arrowhead, out=-8, in =187] (1) to (3);
	\end{pgfonlayer}
\end{scope}

\end{tikzpicture} 
\]

\caption{ Multiple model schemas (boxes are objects, dark arrows are morphisms), connected by functors (the hollow arrows). On the left, DERs with costly transitions share a common underlying base model with a stochastic DER model. On the right, stochastic DER models and stochastic models of natural processes are both modeled using Markov processes.} \label{fig:DERmodels}
\end{figure}

Depending on the analysis to be performed, this basic DER meta-model or schema extends to include additional information such as state of charge, virtual cost of transitions, location, etc. The functorial data model offers a robust collection of ways to translate between such models. Some of these models and the functors relating them are summarized in the following and in Figure \ref{fig:DERmodels}.

In \cite{composableDER1, composableDER2}, virtual costs are assigned to transitions representing the willingness/ability of a controllable DER to perform a certain transition. This leads to a new schema $\DERCost$ with an additional attribute for transitions and obvious inclusion functor to it from the base DER model.

Attaching non-negative rates to transitions on a graph gives a Markov process, summarized by the schema $\Markov$ \cite{norris1998markov}. For now, we implement these rates as `Floats,' tabling the discussion of constraints in CQL until Section \ref{sec:constraints}. Quantifying variability of generation from renewable sources is a key issue when modeling DERs. One approach models DERs as Markov processes, giving a stochastic DER model $\DERMarkov$. 

Over long time scales, the steady state probabilities of such a model can be used to estimate energy production and other performance indices. In \cite{wind}, this approach is utilized to evaluate reliability of small wind farm generation by assigning probabilistic transitions between operative and failed states and coupling this with a stochastic model of wind variability. This methodology is also applied to small hydro electric stations in \cite{water}. Stochastic models of solar irradiance are also used to generate synthetic data for system design \cite{fire}. We summarize these stochastic models of weather in the schema $\Weather$. 

Functors $F \maps M \to N$ between database schemas give rise to adjoint triples of functors $\Sigma_F, \Delta_F, \Pi_F$ between the associated categories of instances, where $\Delta_F \maps N\Inst \to M\Inst$ and $\Sigma_F, \Pi_F \maps M\Inst \to N\Inst$. These functors are related to ``uber-flower'' queries $Q \maps M \to N$. Such a $Q$ can be evaluated (as in other data models) to give a functor $\eval(Q) \maps M\Inst \to N\Inst$ or dually ``coevaluated'' to give a functor $\coeval(Q) \maps N\Inst \to M\Inst$. Using data migration functors and queries, along with CQL's ability to compute colimits of instances, offers a useful way to translate between the data associated to different models.

The models presented here are connected by inclusion functors, summarized in Figure \ref{fig:DERmodels}. Even in the simple setting described here, the use of mappings connecting DER models enables reuse and validation of said models, while providing an extensible framework for model documentation. Implementing mappings between models which utilize the same method, such as Markov processes, enables the reuse of tools or methods, e.g.\ steady state solvers. In Section \ref{sec:connectingtools}, we give a more detailed exposition on how functors between schemas along with their associated adjoint triples and queries can be used to connect models to tools within the paradigm of categorical databases. 

In the next section we give a categorical treatment of DER aggregation, the process of taking collections of DERs and combining them into a single DER. For this we move to the setting of symmetric monoidal categories and consider demand regions as subsets of the complex plane. 

\section{Aggregation as a Symmetric Monoidal Product} \label{sec:DERstructure}
Aggregation of DERs is the key to unlocking their potential to provide ancillary services such as peak shaving and voltage control. It is often third-party aggregators who act as intermediaries between utilities and customers, pooling resources and providing data integration and control strategies via the development of distributed energy resource management systems (DERMS) \cite{NRELagg}. 

In this section, we focus on the basic problem of aggregating demand regions and state spaces of DERs, presenting a symmetric monoidal category $\DER$ whose objects are DERs, and where aggregation serves as the tensor product. This model adds a reflexive property \cite{staymeredith2017opsem}, i.e.\ mandatory self-edges for each node, to the underlying graphs of the DER models outlined previously. Morphisms in $\DER$ correspond to adjusting the level of granularity of the state space.

\begin{defn}
A \textbf{distributed energy resource (DER)} $\D=(S, T, s, t, r, d)$ consists of a graph $s,t \maps T \to S$, together with a function $r \maps S \to T$, satisfying $s \circ r = t \circ r = \id_S$, picking out an \textbf{identity transition} from each state to itself, and a  function $d \maps S \to 2^\CC$ assigning to each state $\sigma \in S$ a \textbf{power demand region} $d(\sigma) \subseteq \CC$.
For each state $\sigma \in S$ we write $1_\sigma = r(\sigma)$ and call $1_\sigma$ the identity transition of $\sigma$.
\end{defn}
This definition is summarized by the diagram $
    \begin{tikzcd}
    T \rar["s" near start, "t" near end, shift left] & S \rar["d"] \lar["r", shift left] & 2^\CC
    \end{tikzcd} $.

\begin{defn}
A \textbf{morphism of DERs} $\phi \maps \D \to \D'$ consists of a pair of functions $(\phi_S, \phi_T)$, where $\phi_S \maps S \to S'$ and $\phi_T \maps T \to T'$, such that
for all $\tau \in T$, $\phi_S(s(\tau)) = s'(\phi_T(\tau))$ and $\phi_S(t(\tau)) = t'(\phi_T(\tau))$, and
for all $\sigma \in S$, $\phi(1_\sigma) = 1_{\phi(\sigma)}$ and $d(\sigma) \subseteq d'(\phi_S(\sigma))$.
Together with these morphisms (and the obvious identity morphisms and composition law), DERs form a category which we denote $\DER$.
\end{defn}

In short, a morphism of DERs is a homomorphism of the underlying graphs that acts as an inclusion of subsets on the demand regions for each state. Such morphisms can be used to translate between models of a DER, e.g.\ by adding more states or by merging states which are indistinguishable in the codomain model. An example of this is provided in Subsection \ref{sec:DERquotient}.

%

Demands can be aggregated using Minkowski sums; see \cite{farouki2001minkowski} for more details as well as \cite{kundu2018approximating} for an application to modeling the flexibility of DERs.
\begin{defn}
Given two subsets $X, Y \subseteq \CC$, the \textbf{Minkowski sum} of $X$ and $Y$ is the set
\[
    X + Y = \{ x + y : (x, y) \in X \times Y \} \subseteq \CC.
\]
Under this operation, $2^\CC$ is a commutative monoid with unit $\{ 0 \}$.
\end{defn}

\begin{defn}
The \textbf{aggregate} of two DERs $\D$ and $\D'$ is the DER $\D \otimes \D' = (S \times S', T \times T', s \times s', t \times t', r \times r', d + d')$, where $d + d' \maps S \times S' \to 2^{\CC}$ is defined by $(d + d')(\sigma, \sigma') = d(\sigma) + d'(\sigma') \subseteq \CC$ for any $(\sigma, \sigma') \in S \times S$. 
\end{defn}

In short, the aggregate of two DERs is the categorical product of the underlying graphs (see \cite{staymeredith2017opsem} or \cite[Proposition~3.3.9]{CTincontext}), where each product state is equipped with demand equal to the Minkowski sum of its factors. Observe that the reflexive property of individual DERs within an aggregate DER enables independent transitions. 

Aggregation extends easily to morphisms by taking Cartesian products of functions, so in this way we see $\otimes \maps \DER \times \DER \to \DER$ is a bifunctor. In fact, letting $\mathcal{I}$ denote the DER with one state $\sigma$, a single transition $1_\sigma$, and power demand $d(\sigma) = \{ 0 \} \subseteq \CC$, it is not hard to show that $\DER$ is a symmetric monoidal category with tensor product $\otimes$ and unit $\mathcal{I}$. As a result, string diagrams can be used to reason about DERs and aggregation \cite{joyal1991geometry, Selinger2011}.


\subsection{Net Demand Quotient} \label{sec:DERquotient}

When aggregating DERs, the state space grows rapidly. For operations at the distribution level, all that is relevant is the \emph{net} power demand. Thus it is natural to mod out by an equivalence relation whereby states with identical power demand are identified. The following definition formalizes this notion.

\begin{defn} 
Let $\D$ be a DER. Consider the equivalence relation $\sim$ on the states $S$ of $\D$ where $\sigma \sim \sigma'$ if and only if $d(\sigma) = d(\sigma')$. This induces an equivalence relation $\approx$ on the edges $T$ of $\D$ where $\tau \approx \tau'$ if and only if $s(\tau) \sim s(\tau')$ and $t(\tau) \sim t(\tau')$. We can define the \textbf{net demand DER} $\overline{\D}$ of $\D$ by $\overline{\D} = (S/{\sim}, T/{\approx}, \overline{s}, \overline{t}, \overline{r}, \overline{d})$, where $\overline{s}, \overline{t}, \overline{r},$ and $\overline{d}$ are defined in the obvious way.
\end{defn}

The equivalence relation above gives rise to a DER morphism $ \overline{ ( \ )} \maps \D \to \overline{\D}$ which identifies states with equal power demand and transitions among them. Composing this morphism with aggregation applied to a pair of DERs $\D$ and $\D'$ gives a DER $\overline{\D \otimes \D'}$ which only distinguishes states which differ in their net power demand.

\begin{figure}[h]

\begin{subfigure}[b]{0.3 \linewidth}
\[ \begin{tikzpicture}[scale=0.6]

\begin{pgfonlayer}{nodelayer}

		\node [style=none] (1) at (0.5, 3.8) {$Q$};
		\node [style=none] (2) at (3, -0.5) {$P$};
		\node [style=none] (3) at (3, -0) {};
		\node [style=none] (4) at (0, 4) {};

		\node [circle, draw=blue, fill=blue] (5) at  (0,0) {};
		\node [circle, draw=blue, fill=blue] (5) at  (2 , 1.5) {};

		\node [style=none] (7) at (-3, -0) {};
		\node [style=none] (8) at (0, -1) {};
	\end{pgfonlayer}
	\begin{pgfonlayer}{edgelayer}
		\draw [style=simple] (4.center) to (8.center);
		\draw [style=simple] (7.center) to (3.center);
	\end{pgfonlayer}

\end{tikzpicture} \]
\caption{HVAC}
\end{subfigure}
\begin{subfigure}[b]{0.3 \linewidth}
\[ 
\begin{tikzpicture}[scale=0.6]

\begin{pgfonlayer}{nodelayer}

		\node [style=none] (1) at (0.5, 3.8) {$Q$};
		\node [style=none] (2) at (3, -0.5) {$P$};
		\node [style=none] (3) at (3, -0) {};
		\node [style=none] (4) at (0, 4) {};

		\node [circle, draw=blue, fill=blue] (5) at  (0,0) {};
		\node [circle, draw=blue, fill=blue] (5) at  (2, 1.5) {};
		\node [circle, draw=blue, fill=blue] (5) at  (-2, 1.5) {};

		\node [style=none] (7) at (-3, -0) {};
		\node [style=none] (8) at (0, -1) {};
	\end{pgfonlayer}
	\begin{pgfonlayer}{edgelayer}
		\draw [style=simple] (4.center) to (8.center);
		\draw [style=simple] (7.center) to (3.center);
	\end{pgfonlayer}

\end{tikzpicture} \]
\caption{Battery}
\end{subfigure}
\begin{subfigure}[b]{0.3 \linewidth}
\[ 
\begin{tikzpicture}[scale=0.6]

\begin{pgfonlayer}{nodelayer}

		\node [style=none] (1) at (0.5, 3.8) {$Q$};
		\node [style=none] (2) at (3.5, -0.5) {$P$};
		\node [style=none] (3) at (3.5, -0) {};
		\node [style=none] (4) at (0, 4) {};

		\node [circle, draw=blue, fill=blue] (5) at  (0,0) {};
		\node [circle, draw=blue, fill=blue] (5) at  (2 , 1.5) {};
		\node [circle, draw=blue, fill=blue] (5) at  (-2, 1.5) {};

		\node [circle, draw=blue, fill=blue] (5) at  (0,3) {};
		\node [circle, draw=blue, fill=blue] (5) at  (4 , 3) {};
		\node [circle, draw=blue, fill=blue] (5) at  (-2, 1.5) {};

		\node [style=none] (7) at (-3.5, -0) {};
		\node [style=none] (8) at (0, -1) {};
	\end{pgfonlayer}
	\begin{pgfonlayer}{edgelayer}
		\draw [style=simple] (4.center) to (8.center);
		\draw [style=simple] (7.center) to (3.center);
	\end{pgfonlayer}

\end{tikzpicture} \]
\caption{HVAC-Battery}
\end{subfigure}

\caption{The demand profile hybrid or aggregate DER consisting of an HVAC system and a battery.}
\end{figure}
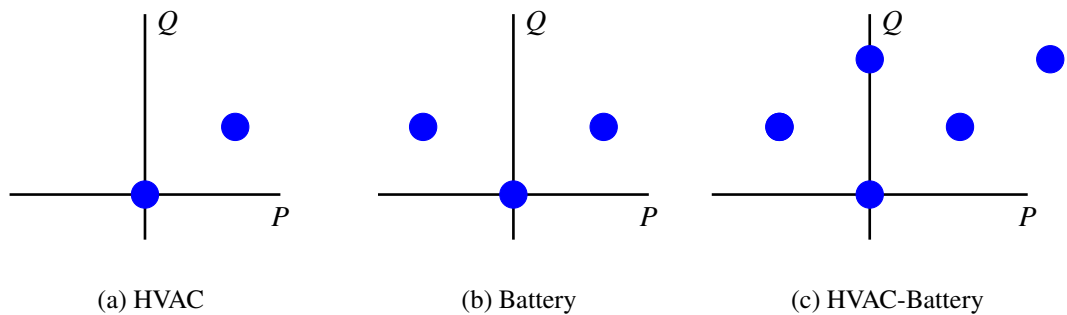

Any path in $\overline{\D}$ will give a set of paths in $\D$ traveling among DER states. We can then consider methods for selecting the `best' or `least-costly' sequence of DER transitions which accomplish some desired transition in net demand. This allows for dynamic tasking of DERs to accommodate demand fluctuations without requiring distribution level operators to have full knowledge of the details of a collection of DERs.


DER aggregation is typically done locally/regionally, interfacing with grid operators at the distribution or transmission level where the problem becomes matching generation with consumption while maintaining stable operating conditions. We now turn to the basic problem of distributing electricity through the grid so as to match production and consumption. specification and numerical solution of basic power flow problems using categorical databases.


\section{Power Flow Problems} \label{sec:PF}

In this section, we show how to connect models with tools or solvers by describing the specification and numerical solution of basic power flow problems using categorical databases. This amounts to finding solutions to a set of non-linear equations, the power flow equations, defined over a network or power flow graph:


\begin{defn}
A \textbf{power flow graph} consists of graph $s,t \maps E \to N$, together with functions $g, b \maps E \to \R$, assigning a \textbf{conductance} and \textbf{susceptance} to each edge. Nodes in the graph $n \in N$ are typically called \textbf{buses}, while edges $e \in E$ are referred to as \textbf{branches}. Conductance and susceptance are the real and imaginary parts of the complex admittance, a measure of the susceptibility of a branch to admitting current flow. 
\end{defn}

\begin{figure}[h]
\[
 \begin{tikzpicture}[scale=.75, font=\tiny]
	\begin{pgfonlayer}{nodelayer}
		\node [style=wbox, inner sep = 3pt] (0) at (0, 2) {$PQ$ Bus};
		\node [style=wbox, inner sep = 3pt] (1) at (-4, -0) {Branch};
		\node [style=wbox, inner sep = 3pt] (2) at (3.1, -0) {Generator};
		\node [style=wbox, inner sep = 3pt] (3) at (0, -0) {Bus};
		\node [style=wbox, inner sep = 3pt] (4) at (0, -2) {$PV$ Bus};
		\node [style=none] (5) at (-2, 1) {source};
		\node [style=none] (6) at (-2, -0.25) {target};
		\node [style=none] (7) at (0.75, 1) {};
		\node [style=none] (8) at (1.5, 0.25) {};
		\node [style=none] (9) at (-1, -1.25) {};
		\node [style=none] (10) at (1.75, -1.25) {};
			\end{pgfonlayer}
	\begin{pgfonlayer}{edgelayer}
		\draw [style=arrowhead, bend left, looseness=1.00] (1) to (3);
		\draw [style=arrowhead, bend right, looseness=1.00] (1) to (3);
		\draw [style=arrowhead] (0) to (3);
		\draw [style=arrowhead] (2) to (3);
		\draw [style=arrowhead] (4) to (2);
		\draw [style=arrowhead] (4) to (3);
	\end{pgfonlayer}
\end{tikzpicture} 
\]
\caption{A schema describing a generic power flow problem. A $PQ$ bus represents a typical load, whose real and reactive power demands are known and fixed, at any moment of time. All $PV$ buses are viewed as having generators attached, producing constant power at a specific voltage. Slack buses are omitted for visual clarity.} 
\label{fig:PowerFlow}
\end{figure}
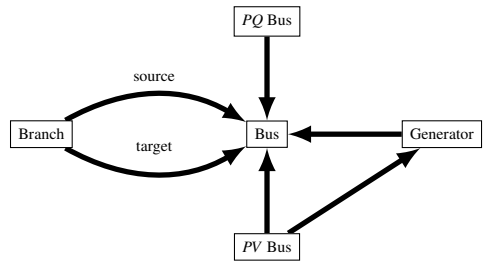

The variables of interest are the real and imaginary parts of the complex power $P+iQ$ and the magnitude and phase of the complex voltage $Ve^{i\theta}$, which we regard as partial functions $P,Q,V,\theta \maps N \to \R$. Buses are typed as $PQ$, $PV$, or $V\theta$ buses according to which pair of variables is regarded as fixed, see Figure \ref{fig:PowerFlow}. The remaining free variables are determined by solving the power balance equations.


\begin{defn}
The \textbf{power balance equations} \cite{kundur} for a power flow graph are the $2 |N|$ equations
\[
\begin{split}
    P_i &= V_i \sum_j V_j \left( \ g_{ij} \cos(\theta_i - \theta_j) + b_{ij} \sin(\theta_i - \theta_j) \ \right) \\
    Q_i &= V_i \sum_j V_j \left(\  g_{ij} \sin(\theta_i - \theta_j) - b_{ij} \cos(\theta_i - \theta_j) \ \right),
\end{split}
\]
where we write $P_i := P(N_i)$ and $g_{ij} := g(E_{ij})$ etc.\ and each sum is taken over all buses adjacent to $i$.
\end{defn}

We summarize the data needed to specify a power flow problem in a CQL schema in Figure \ref{fig:PowerFlow}, omitting attributes for simplicity.

Due in part to their non-linearity, solving the power flow equations is typically done numerically either using freely available software, commercial tools, or customized code. Such tools usually require specific solver parameters and use their own internal data structures. 

\subsection{Connecting to a Tool} \label{sec:toolint}

\begin{figure}[h]
    \centering
    \begin{tikzpicture}[scale=0.65, font=\tiny]
    \begin{scope}[shift={(12,0)}]
        \begin{pgfonlayer}{nodelayer}
            \node [style=wbox] (0) at (0, 0) {\begin{tabular}{c} Parameters \\ \hline Algorithm : String \\ Max\_Iterations : Int \\ Tolerance : Float \\ \end{tabular}};
        \end{pgfonlayer}
    \end{scope} 
   
	\begin{pgfonlayer}{nodelayer}
		\node [style=wbox] (1) at (-5.4, -0) {\tiny \begin{tabular}{c} Branch \\ \hline F\_BUS :Int \\ T\_BUS : Int \end{tabular}};
		\node [style=wbox] (2) at (5, -0) {\tiny \begin{tabular}{c} Generator \\ \hline BUS : Int \\ $P_G$ : Float \\ $V_G$ : Float \end{tabular}};
		\node [style=wbox] (3) at (0, -0) {\tiny \begin{tabular}{c} Bus \\ \hline BUS\_I : Int \\ BUS\_TYPE : Int \\ $P_D$ : Float \\ $Q_D$ : Float \\ $V_M$ : Float \\ $V_A$ : Float \\ \end{tabular}};
	\end{pgfonlayer}
	
	\begin{pgfonlayer}{edgelayer}
		\draw [style=arrowhead, out=8, in=175] (1) to (3);
		\draw [style=arrowhead, out=-8, in =185] (1) to (3);
		\draw [style=arrowhead] (2) to (3);
	\end{pgfonlayer}
\end{tikzpicture} 
	
    \caption{A MATPOWER power flow schema on the left, with solver parameters on the right. For simplicity we only show a few attributes for each entity. Attribute names are based on those in MATPOWER. Compare Figure \ref{fig:PowerFlow}, which was developed based on a reorganization of the schema on the left.}
    \label{fig:easik}
\end{figure}
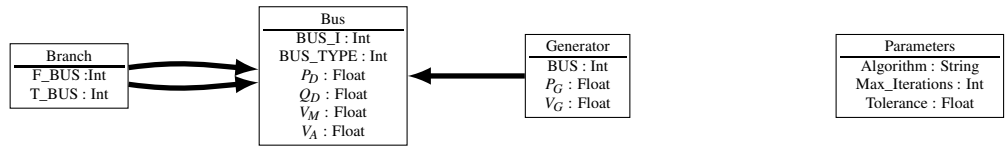

MATPOWER is a commonly used power systems toolbox, implemented in MATLAB. The MATPOWER data format specifications are organized into tables in Appendix B of the MATPOWER manual \cite{matpower}. We translate these specifications into MATPOWER-specific schemas in CQL. Figure \ref{fig:easik} shows the resulting schemas representing both a power flow problem as well as an associated solver, e.g\ We characterize an iterative Newton-Raphson solver in terms of its required parameters such as tolerance, maximum number of iterations, etc. 

Encoding the input problem specification, the output solution structure, as well as the solver parameters in database schemas enables systematic experimentation, i.e.\ varying inputs or parameters, while providing flexible and traceable documentation, i.e.\ storing just solutions or including the solver settings used in each run. The input and output features common to all solvers of a given type can be organized into a generic schema for solvers of that given type.

\subsection{Constraints in CQL}\label{sec:constraints}
CQL allows for the enforcement of constraints in the form of path equations. For example, consider the chunk of our MATPOWER schema:
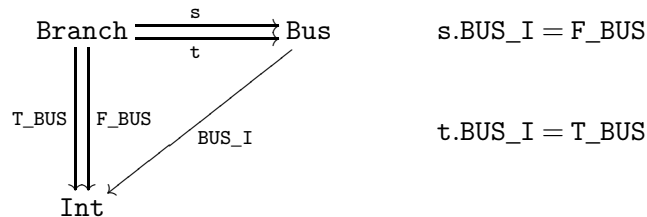
\[ \begin{array}{ccc} \xymatrix{ \mathtt{Branch}  \ar@<0.5ex>[rr]^{\mathtt{s}} \ar@<-0.5ex>[rr]_{\mathtt{t}}  \ar@<0.5ex>[dd]^{\mathtt{F\_BUS}}  \ar@<-0.5ex>[dd]_{\mathtt{T\_BUS}} &  & \mathtt{Bus} \ar@<0.3ex>[ddll]^{\mathtt{BUS\_I} } \\ \\  \mathtt{Int} & & } & \ \ \ \ \ & \xymatrix{ \mathtt{s.BUS\_I = F\_BUS} \\ \mathtt{ t.BUS\_I = T\_BUS} } \end{array}\]
The equations on the right enforce the constraint that the indexing of buses via $\mathtt{BUS\_I}$ is consistent with the indexing of $\mathtt{T\_BUS}$ and $\mathtt{F\_BUS}$ of branches.


\subsection{Connecting to DERs} \label{sec:PFplusDER}



\begin{figure}[h]
    \centering
    \[
    \begin{tikzpicture}[scale=0.8]
        \begin{pgfonlayer}{nodelayer}
            \node [style=wbox, inner sep = 2.5pt ] (0) at (-5, 0) {Bus 1};
            \node [style=wbox, inner sep = 2.5pt] (1) at (-5, 1) {Bus 2};
            \node [style=wbox, inner sep = 2.5pt] (2) at (-4, -1) {Bus 3};
            \node [style=wbox, inner sep = 2.5pt] (3) at (-6, -1) {Bus 4};
            \node [style=wbox] (4) at (0, -0.25) {\begin{tabular}{c} $P = 2.0$ \\ $Q = 5.0$ \end{tabular}};
            \node [style=wbox] (5) at (4, 0.5) {\begin{tabular}{c} 5 Batteries \\ 3 HVAC \\ 1 Photovoltaic \end{tabular}};
        \end{pgfonlayer}
        \begin{pgfonlayer}{edgelayer}
            \draw (0) to (1);
            \draw (0) to (2);
            \draw (0) to (3);
            \draw (2.north east) to (4.north west);
            \draw (2.south east) to (4.south east);
            \draw (4.north west) to (5.north west);
            \draw (4.south east) to (5.south east);
        \end{pgfonlayer}
    \end{tikzpicture}
    \]
    \caption{Incorporating data from a collection of DERs into a node in a power flow graph. Units and values of $P$, $Q$ are arbitrary.}
    \label{fig:multiscale}
\end{figure}
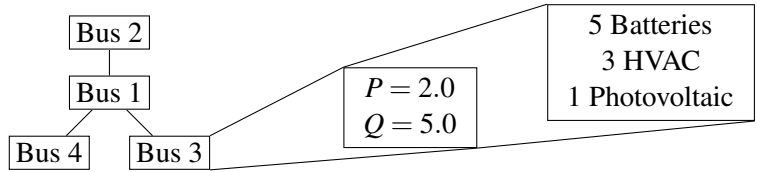

To interface with a standard power flow problem, we place DERs at the relevant nodes of a power flow graph, treating each such node as a $PQ$ bus. For each such bus we determine average $P$, $Q$ values from the DERs at that node, for example by modeling the relevant DERs as Markov chains, as described in Subsection \ref{sec:modeltranslation}, and returning the sum of the expected steady-state $P$, $Q$ values for each DER. This process is depicted in Figure \ref{fig:multiscale} and implemented by the authors in a MATPOWER example.

This hybrid setup enables the exploration on the dependence of overall solutions to the power flow equations on the types and behaviors of DERs, e.g.\ how stochasticity of distributed generation enters into overall power distribution. We now turn to the problem of connecting multiple tools or solvers.

\subsection{Connecting Tools} \label{sec:connectingtools}
Modeling something as complex as the electrical grid typically involves collaborations among teams who may utilize a varLy of tools or implementations, even for the same or similar problems. This creates a need for translation  and validation among different solvers. We describe a procedure for accomplishing this task using techniques from the functorial data model, as presented in Subsection \ref{sec:modeltranslation}. 

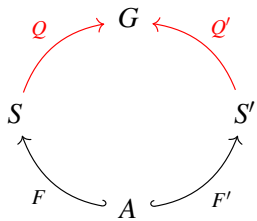
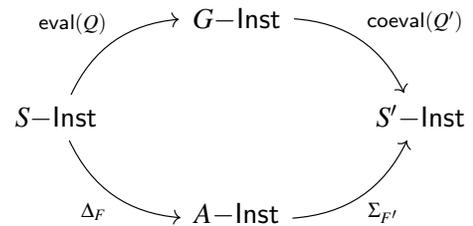
\begin{figure}[h]
    \centering
    \begin{subfigure}[b]{.4\linewidth}
        \centering
        \begin{tikzcd}
            & G \\
            S \urar["Q", bend left, red] & & S' \ular["Q'", bend right, swap, red] \\
            & A \ular["F", bend left, hook'] \urar["F'", bend right, hook, swap]
        \end{tikzcd}
        \caption{Transformation between different solver schemas for the same problem.}
        \label{fig:schematransformation}
    \end{subfigure} \hfill
    \begin{subfigure}[b]{.4\linewidth}
        \centering
        \begin{tikzcd}
            & G\Inst \drar["\coeval(Q')", bend left] \\
            S\Inst \urar["\eval(Q)", bend left] \drar["\Delta_F", bend right, swap] & & S'\Inst \\
            & A\Inst \urar["\Sigma_{F'}", bend right, swap]
        \end{tikzcd}
        \caption{Transformation between instances of different solver schemas.}
        \label{fig:insttransformation}
    \end{subfigure}
    
    \caption{Diagram depicting the transformation of instances for solver schema $S$ to instances for solver schema $S'$. Black arrows are functors; red arrows are queries.}
    \label{fig:solvertransformation}
\end{figure}

Figure \ref{fig:solvertransformation} provides diagrams describing how to translate between solvers. Consider two solvers for some problem, represented by schemas $S$ and $S'$, e.g.\ the schema in Figure \ref{fig:easik} and a schema for a solver with a different set of parameters. One can construct a generic solver schema $G$ for the problem, e.g.\ that in Figure \ref{fig:PowerFlow}, along with queries $Q \maps S \to G$ and $Q' \maps S' \to G$, specifying which information is shared among the generic and specific instances. In this case, one should also define an auxiliary schema $A$ for data which appears in both $S$ and $S'$ but not in $G$, as well as functors $F \maps A \to S$ and $F' \maps A \to S'$ inserting the data of $A$ into both specific solver schemas.

These constructions give rise to functors between the associated categories of instances, as depicted in Figure \ref{fig:insttransformation}. For every instance $I$ of $S$, we can obtain two instances of $S'$ by applying these functors. These can be combined using a suitable colimit in $S'\Inst$ to get a single instance of $S'$ containing all possible data from $S$. Such a construction enables one to translate between the inputs, outputs, and parameters for the solver represented by $S$ and the corresponding values for the solver represented by $S'$.


\section{Conclusions and Future Work}
This paper provides a window into our efforts to concretize the potential utility of a category theoretic viewpoint for problems dealing with multiple related models and tools in the context of power systems engineering. We saw that techniques and tools from categorical databases can readily be applied to specify and translate among various models, connecting those models to particular analysis tools, as well as connecting various tools themselves. 

Further work is required to extend this category-theoretic modeling paradigm to other engineering domains as well as within power systems. What is desired is not a modeling framework which captures the full complexity of the today's grid, but rather a framework which enables the expedient exploration and evaluation of various possible future architectures and pathways to those. The need for such a modeling ecosystem is not unique to power systems. 

Of particular relevance for future work in Smart Grid technologies are aspects of control and communication enabled by new devices such as Smart Meters and increased deployment of phasor measurement units (PMUs), devices which measure current, voltage, or phase across the grid. Managing this coupling of an information network with a physical power network presents ample opportunities for applied category theorists. 

Lastly, further development of tools for specifying and modeling systems using category theory, e.g.\ CQL, is essential in terms of engagement with domains. Being able to point practitioners to a system they can get their hands on and play with goes a long way towards arriving at a useful common understanding.



\vskip 1em

{ \small \textbf{Acknowledgements} Source code for the examples discussed can be found at \href{https://github.com/isabit/AQL_Powersystems}{github:AQL\_Powersystems}. The authors would like to thank Prof.\ Mads Almassalkhi and his group at the University of Vermont for sharing their knowledge of power systems, as well as David Spivak and Ryan Wisnesky of Conxeus AI for helpful discussions. This material is based upon work supported by the National Science Foundation under Grant No.1746077. Any opinions, findings, and conclusions or recommendations expressed in this material are those of the author(s) and do not necessarily reflect the views of the National Science Foundation. JN was supported by the NIST SURF and NIST PREP programs. BP was supported by an NRC Postdoctoral Research Associateship. }

{\small {\bf Official contribution of the National Institute of Standards and Technology;} not subject to copyright in the United States. Certain commercial equipment, instruments, or materials are identified in this paper in order to specify the experimental procedure adequately. Such identification is not intended to imply recommendation or endorsement by the National Institute of Standards and Technology, nor is it intended to imply that the materials or equipment identified are necessarily the best available for the purpose. Parts of this paper may have been presented in technical seminars and included in government publications. Recorded versions of those seminars and copyright free versions of publications are available through the National Institute of Standards and Technology.
}

\nocite{*}
\bibliographystyle{eptcs}
\bibliography{bibliography}

\begin{thebibliography}{10}
\providecommand{\bibitemdeclare}[2]{}
\providecommand{\surnamestart}{}
\providecommand{\surnameend}{}
\providecommand{\urlprefix}{Available at }
\providecommand{\url}[1]{\texttt{#1}}
\providecommand{\href}[2]{\texttt{#2}}
\providecommand{\urlalt}[2]{\href{#1}{#2}}
\providecommand{\doi}[1]{doi:\urlalt{http://dx.doi.org/#1}{#1}}
\providecommand{\bibinfo}[2]{#2}

\bibitemdeclare{article}{baez2015categories}
\bibitem{baez2015categories}
\bibinfo{author}{John~C \surnamestart Baez\surnameend} \&
  \bibinfo{author}{Jason \surnamestart Erbele\surnameend}
  (\bibinfo{year}{2015}): \emph{\bibinfo{title}{Categories in control}}.
\newblock {\sl \bibinfo{journal}{Theory and Applications of Categories}}
  \bibinfo{volume}{30}(\bibinfo{number}{24}), pp. \bibinfo{pages}{836--881}.
\newblock \urlprefix\url{http://www.tac.mta.ca/tac/volumes/30/24/30-24.pdf}.

\bibitemdeclare{article}{baez2018compositional}
\bibitem{baez2018compositional}
\bibinfo{author}{John~C \surnamestart Baez\surnameend} \&
  \bibinfo{author}{Brendan \surnamestart Fong\surnameend}
  (\bibinfo{year}{2018}): \emph{\bibinfo{title}{A compositional framework for
  passive linear networks}}.
\newblock {\sl \bibinfo{journal}{Theory and Applications of Categories}}
  \bibinfo{volume}{33}(\bibinfo{number}{38}), pp. \bibinfo{pages}{1158--1222}.
\newblock \urlprefix\url{http://www.tac.mta.ca/tac/volumes/33/38/33-38.pdf}.

\bibitemdeclare{article}{baez2018open}
\bibitem{baez2018open}
\bibinfo{author}{John~C \surnamestart Baez\surnameend} \& \bibinfo{author}{Jade
  \surnamestart Master\surnameend} (\bibinfo{year}{2018}):
  \emph{\bibinfo{title}{Open {Petri} Nets}}.
\newblock {\sl \bibinfo{journal}{arXiv preprint arXiv:1808.05415}}.

\bibitemdeclare{article}{baez2017compositional}
\bibitem{baez2017compositional}
\bibinfo{author}{John~C \surnamestart Baez\surnameend} \&
  \bibinfo{author}{Blake~S \surnamestart Pollard\surnameend}
  (\bibinfo{year}{2017}): \emph{\bibinfo{title}{A compositional framework for
  reaction networks}}.
\newblock {\sl \bibinfo{journal}{Reviews in Mathematical Physics}}
  \bibinfo{volume}{29}(\bibinfo{number}{09}), p. \bibinfo{pages}{1750028},
  \doi{10.1142/s0129055x17500283}.

\bibitemdeclare{book}{bjg2007digraphs}
\bibitem{bjg2007digraphs}
\bibinfo{author}{J{\o}rgen \surnamestart Bang-Jensen\surnameend} \&
  \bibinfo{author}{Gregory \surnamestart Gutin\surnameend}
  (\bibinfo{year}{2009}): \emph{\bibinfo{title}{Digraphs: Theory, Algorithms
  and Applications}}, \bibinfo{edition}{second} edition.
\newblock \bibinfo{series}{Springer Monographs in Mathematics},
  \bibinfo{publisher}{Springer London}, \doi{10.1007/978-1-84800-998-1}.

\bibitemdeclare{article}{composableDER1}
\bibitem{composableDER1}
\bibinfo{author}{Andrey \surnamestart Bernstein\surnameend},
  \bibinfo{author}{Lorenzo \surnamestart Reyes-Chamorro\surnameend},
  \bibinfo{author}{Jean-Yves~Le \surnamestart Boudec\surnameend} \&
  \bibinfo{author}{Mario \surnamestart Paolone\surnameend}
  (\bibinfo{year}{2015}): \emph{\bibinfo{title}{A composable method for
  real-time control of active distribution networks with explicit power
  setpoints. Part {I}: Framework}}.
\newblock {\sl \bibinfo{journal}{Electric Power Systems Research}}
  \bibinfo{volume}{125}, pp. \bibinfo{pages}{254 -- 264},
  \doi{10.1016/j.epsr.2015.03.023}.

\bibitemdeclare{article}{water}
\bibitem{water}
\bibinfo{author}{C.~L.~T. \surnamestart {Borges}\surnameend} \&
  \bibinfo{author}{R.~J. \surnamestart {Pinto}\surnameend}
  (\bibinfo{year}{2008}): \emph{\bibinfo{title}{Small Hydro Power Plants Energy
  Availability Modeling for Generation Reliability Evaluation}}.
\newblock {\sl \bibinfo{journal}{IEEE Transactions on Power Systems}}
  \bibinfo{volume}{23}(\bibinfo{number}{3}), pp. \bibinfo{pages}{1125--1135},
  \doi{10.1109/TPWRS.2008.926713}.

\bibitemdeclare{article}{stochastic}
\bibitem{stochastic}
\bibinfo{author}{Carmen Lucia~Tancredo \surnamestart Borges\surnameend}
  (\bibinfo{year}{2012}): \emph{\bibinfo{title}{An overview of reliability
  models and methods for distribution systems with renewable energy distributed
  generation}}.
\newblock {\sl \bibinfo{journal}{Renewable and Sustainable Energy Reviews}}
  \bibinfo{volume}{16}(\bibinfo{number}{6}), pp. \bibinfo{pages}{4008 -- 4015},
  \doi{10.1016/j.rser.2012.03.055}.

\bibitemdeclare{article}{functorial}
\bibitem{functorial}
\bibinfo{author}{Spencer \surnamestart Breiner\surnameend},
  \bibinfo{author}{Blake \surnamestart Pollard\surnameend} \&
  \bibinfo{author}{Eswaran \surnamestart Subrahmanian\surnameend}
  (\bibinfo{year}{2019}): \emph{\bibinfo{title}{Functorial Model Management}}.
\newblock {\sl \bibinfo{journal}{Proceedings of the Design Society:
  International Conference on Engineering Design}}
  \bibinfo{volume}{1}(\bibinfo{number}{1}), pp. \bibinfo{pages}{1963--1972},
  \doi{10.1017/dsi.2019.202}.

\bibitemdeclare{article}{capitanescu2016critical}
\bibitem{capitanescu2016critical}
\bibinfo{author}{Florin \surnamestart Capitanescu\surnameend}
  (\bibinfo{year}{2016}): \emph{\bibinfo{title}{Critical review of recent
  advances and further developments needed in {AC} optimal power flow}}.
\newblock {\sl \bibinfo{journal}{Electric Power Systems Research}}
  \bibinfo{volume}{136}, pp. \bibinfo{pages}{57--68},
  \doi{10.1016/j.epsr.2016.02.008}.

\bibitemdeclare{inproceedings}{chalkiadakis2011cooperatives}
\bibitem{chalkiadakis2011cooperatives}
\bibinfo{author}{Georgios \surnamestart Chalkiadakis\surnameend},
  \bibinfo{author}{Valentin \surnamestart Robu\surnameend},
  \bibinfo{author}{Ramachandra \surnamestart Kota\surnameend},
  \bibinfo{author}{Alex \surnamestart Rogers\surnameend} \&
  \bibinfo{author}{Nicholas~R \surnamestart Jennings\surnameend}
  (\bibinfo{year}{2011}): \emph{\bibinfo{title}{Cooperatives of distributed
  energy resources for efficient virtual power plants}}.
\newblock In: {\sl \bibinfo{booktitle}{The 10th International Conference on
  Autonomous Agents and Multiagent Systems-Volume 2}},
  \bibinfo{organization}{International Foundation for Autonomous Agents and
  Multiagent Systems}, pp. \bibinfo{pages}{787--794},
  \doi{10.5555/2031678.2031730}.

\bibitemdeclare{inproceedings}{gridlabd1}
\bibitem{gridlabd1}
\bibinfo{author}{D.~P. \surnamestart Chassin\surnameend},
  \bibinfo{author}{K.~\surnamestart Schneider\surnameend} \&
  \bibinfo{author}{C.~\surnamestart Gerkensmeyer\surnameend}
  (\bibinfo{year}{2008}): \emph{\bibinfo{title}{{GridLAB-D}: An open-source
  power systems modeling and simulation environment}}.
\newblock In: {\sl \bibinfo{booktitle}{2008 {IEEE}/{PES} Transmission and
  Distribution Conference and Exposition}}, \bibinfo{publisher}{{IEEE}},
  \doi{10.1109/tdc.2008.4517260}.

\bibitemdeclare{article}{gridlabd2}
\bibitem{gridlabd2}
\bibinfo{author}{David~P. \surnamestart Chassin\surnameend},
  \bibinfo{author}{Jason~C. \surnamestart Fuller\surnameend} \&
  \bibinfo{author}{Ned \surnamestart Djilali\surnameend}
  (\bibinfo{year}{2014}): \emph{\bibinfo{title}{{GridLAB-D}: An Agent-Based
  Simulation Framework for Smart Grids}}.
\newblock {\sl \bibinfo{journal}{Journal of Applied Mathematics}}
  \bibinfo{volume}{2014}, pp. \bibinfo{pages}{1--12},
  \doi{10.1155/2014/492320}.

\bibitemdeclare{techreport}{NRELagg}
\bibitem{NRELagg}
\bibinfo{author}{Jeffrey~J. \surnamestart Cook\surnameend},
  \bibinfo{author}{Kristen \surnamestart Ardani\surnameend},
  \bibinfo{author}{Eric \surnamestart O’Shaughnessy\surnameend},
  \bibinfo{author}{Brittany \surnamestart Smith\surnameend} \&
  \bibinfo{author}{Robert \surnamestart Margolis\surnameend}
  (\bibinfo{year}{2019}): \emph{\bibinfo{title}{Expanding PV Value: Lessons
  Learned from Utility-led Distributed Energy Resource Aggregation in the
  United States}}.
\newblock \bibinfo{type}{Technical Report}, \bibinfo{address}{Golden, CO},
  \doi{10.2172/1483067}.

\bibitemdeclare{book}{clrs2009algorithms}
\bibitem{clrs2009algorithms}
\bibinfo{author}{Thomas~H. \surnamestart Cormen\surnameend},
  \bibinfo{author}{Charles~E. \surnamestart Leiserson\surnameend},
  \bibinfo{author}{Ronald~L. \surnamestart Rivest\surnameend} \&
  \bibinfo{author}{Clifford \surnamestart Stein\surnameend}
  (\bibinfo{year}{2009}): \emph{\bibinfo{title}{Introduction to Algorithms,
  Third Edition}}, \bibinfo{edition}{3rd} edition.
\newblock \bibinfo{publisher}{The MIT Press}.

\bibitemdeclare{article}{courser2017bicategory}
\bibitem{courser2017bicategory}
\bibinfo{author}{Kenny \surnamestart Courser\surnameend}
  (\bibinfo{year}{2017}): \emph{\bibinfo{title}{A bicategory of decorated
  cospans}}.
\newblock {\sl \bibinfo{journal}{Theory and Applications of Categories}}
  \bibinfo{volume}{32}(\bibinfo{number}{995}), p. \bibinfo{pages}{1027}.
\newblock \urlprefix\url{http://www.tac.mta.ca/tac/volumes/32/29/32-29.pdf}.

\bibitemdeclare{article}{diskin2014category}
\bibitem{diskin2014category}
\bibinfo{author}{Zinovy \surnamestart Diskin\surnameend} \&
  \bibinfo{author}{Tom \surnamestart Maibaum\surnameend}
  (\bibinfo{year}{2014}): \emph{\bibinfo{title}{Category theory and
  model-driven engineering: From formal semantics to design patterns and
  beyond}}.
\newblock {\sl \bibinfo{journal}{Model-Driven Engineering of Information
  Systems: Principles, Techniques, and Practice}}, p. \bibinfo{pages}{173},
  \doi{10.1201/b17480-7}.

\bibitemdeclare{article}{farouki2001minkowski}
\bibitem{farouki2001minkowski}
\bibinfo{author}{Rida~T \surnamestart Farouki\surnameend},
  \bibinfo{author}{Hwan~Pyo \surnamestart Moon\surnameend} \&
  \bibinfo{author}{Bahram \surnamestart Ravani\surnameend}
  (\bibinfo{year}{2001}): \emph{\bibinfo{title}{Minkowski geometric algebra of
  complex sets}}.
\newblock {\sl \bibinfo{journal}{Geometriae Dedicata}}
  \bibinfo{volume}{85}(\bibinfo{number}{1-3}), pp. \bibinfo{pages}{283--315},
  \doi{10.1023/A:1010318011860}.

\bibitemdeclare{book}{Feller}
\bibitem{Feller}
\bibinfo{author}{William \surnamestart Feller\surnameend}
  (\bibinfo{year}{1957}): \emph{\bibinfo{title}{An Introduction to Probability
  Theory and its Applications}}, \bibinfo{edition}{2} edition.
\newblock \bibinfo{volume}{1}, \bibinfo{publisher}{John Wiley \& Sons, Inc.}

\bibitemdeclare{article}{fleming2003database}
\bibitem{fleming2003database}
\bibinfo{author}{Michael \surnamestart Fleming\surnameend},
  \bibinfo{author}{Ryan \surnamestart Gunther\surnameend} \&
  \bibinfo{author}{Robert \surnamestart Rosebrugh\surnameend}
  (\bibinfo{year}{2003}): \emph{\bibinfo{title}{A database of categories}}.
\newblock {\sl \bibinfo{journal}{Journal of Symbolic Computation}}
  \bibinfo{volume}{35}(\bibinfo{number}{2}), pp. \bibinfo{pages}{127 -- 135},
  \doi{10.1016/S0747-7171(02)00104-9}.

\bibitemdeclare{article}{fong2015decorated}
\bibitem{fong2015decorated}
\bibinfo{author}{Brendan \surnamestart Fong\surnameend} (\bibinfo{year}{2015}):
  \emph{\bibinfo{title}{Decorated cospans}}.
\newblock {\sl \bibinfo{journal}{Theory and Applications of Categories}}
  \bibinfo{volume}{30}(\bibinfo{number}{33}), pp. \bibinfo{pages}{1096--1120}.
\newblock \urlprefix\url{http://www.tac.mta.ca/tac/volumes/30/33/30-33.pdf}.

\bibitemdeclare{article}{johnson2002entity}
\bibitem{johnson2002entity}
\bibinfo{author}{Michael \surnamestart Johnson\surnameend},
  \bibinfo{author}{Robert \surnamestart Rosebrugh\surnameend} \&
  \bibinfo{author}{RJ~\surnamestart Wood\surnameend} (\bibinfo{year}{2002}):
  \emph{\bibinfo{title}{Entity-relationship-attribute designs and sketches}}.
\newblock {\sl \bibinfo{journal}{Theory and Applications of Categories}}
  \bibinfo{volume}{10}(\bibinfo{number}{3}), pp. \bibinfo{pages}{94--112}.
\newblock \urlprefix\url{http://www.tac.mta.ca/tac/volumes/10/3/10-03.pdf}.

\bibitemdeclare{article}{joyal1991geometry}
\bibitem{joyal1991geometry}
\bibinfo{author}{André \surnamestart Joyal\surnameend} \&
  \bibinfo{author}{Ross \surnamestart Street\surnameend}
  (\bibinfo{year}{1991}): \emph{\bibinfo{title}{The geometry of tensor
  calculus, {I}}}.
\newblock {\sl \bibinfo{journal}{Advances in Mathematics}}
  \bibinfo{volume}{88}(\bibinfo{number}{1}), pp. \bibinfo{pages}{55 -- 112},
  \doi{10.1016/0001-8708(91)90003-P}.

\bibitemdeclare{book}{FrankKelly}
\bibitem{FrankKelly}
\bibinfo{author}{F.~P. \surnamestart Kelly\surnameend} (\bibinfo{year}{2011}):
  \emph{\bibinfo{title}{Reversibility and Stochastic Networks}}.
\newblock \bibinfo{publisher}{Cambridge University Press},
  \bibinfo{address}{New York, NY, USA}.

\bibitemdeclare{inproceedings}{kundu2018approximating}
\bibitem{kundu2018approximating}
\bibinfo{author}{Soumya \surnamestart Kundu\surnameend},
  \bibinfo{author}{Karanjit \surnamestart Kalsi\surnameend} \&
  \bibinfo{author}{Scott \surnamestart Backhaus\surnameend}
  (\bibinfo{year}{2018}): \emph{\bibinfo{title}{Approximating flexibility in
  distributed energy resources: A geometric approach}}.
\newblock In: {\sl \bibinfo{booktitle}{2018 Power Systems Computation
  Conference (PSCC)}}, \bibinfo{organization}{IEEE}, pp. \bibinfo{pages}{1--7},
  \doi{10.23919/pscc.2018.8442600}.

\bibitemdeclare{book}{kundur}
\bibitem{kundur}
\bibinfo{author}{Prabha \surnamestart Kundur\surnameend}
  (\bibinfo{year}{1994}): \emph{\bibinfo{title}{Power System Stability and
  Control}}.
\newblock \bibinfo{publisher}{McGraw-Hill}.

\bibitemdeclare{incollection}{CWMmonoids}
\bibitem{CWMmonoids}
\bibinfo{author}{Saunders~Mac \surnamestart Lane\surnameend}
  (\bibinfo{year}{1978}): \emph{\bibinfo{title}{Monoids}}.
\newblock In: {\sl \bibinfo{booktitle}{Categories for the Working
  Mathematician}}, \bibinfo{publisher}{Springer New York}, pp.
  \bibinfo{pages}{161--190}, \doi{10.1007/978-1-4757-4721-8_8}.

\bibitemdeclare{article}{wind}
\bibitem{wind}
\bibinfo{author}{A.~P. \surnamestart {Leite}\surnameend},
  \bibinfo{author}{C.~L.~T. \surnamestart {Borges}\surnameend} \&
  \bibinfo{author}{D.~M. \surnamestart {Falcao}\surnameend}
  (\bibinfo{year}{2006}): \emph{\bibinfo{title}{Probabilistic Wind Farms
  Generation Model for Reliability Studies Applied to Brazilian Sites}}.
\newblock {\sl \bibinfo{journal}{IEEE Transactions on Power Systems}}
  \bibinfo{volume}{21}(\bibinfo{number}{4}), pp. \bibinfo{pages}{1493--1501},
  \doi{10.1109/TPWRS.2006.881160}.

\bibitemdeclare{article}{mens2006taxonomy}
\bibitem{mens2006taxonomy}
\bibinfo{author}{Tom \surnamestart Mens\surnameend} \&
  \bibinfo{author}{Pieter~Van \surnamestart Gorp\surnameend}
  (\bibinfo{year}{2006}): \emph{\bibinfo{title}{A Taxonomy of Model
  Transformation}}.
\newblock {\sl \bibinfo{journal}{Electronic Notes in Theoretical Computer
  Science}} \bibinfo{volume}{152}, pp. \bibinfo{pages}{125--142},
  \doi{10.1016/j.entcs.2005.10.021}.

\bibitemdeclare{article}{momoh1999review}
\bibitem{momoh1999review}
\bibinfo{author}{James~A \surnamestart Momoh\surnameend},
  \bibinfo{author}{Rambabu \surnamestart Adapa\surnameend} \&
  \bibinfo{author}{ME~\surnamestart El-Hawary\surnameend}
  (\bibinfo{year}{1999}): \emph{\bibinfo{title}{A review of selected optimal
  power flow literature to 1993. I. {Nonlinear} and quadratic programming
  approaches}}.
\newblock {\sl \bibinfo{journal}{IEEE transactions on power systems}}
  \bibinfo{volume}{14}(\bibinfo{number}{1}), pp. \bibinfo{pages}{96--104},
  \doi{10.1109/59.744492}.

\bibitemdeclare{article}{momoh1999review2}
\bibitem{momoh1999review2}
\bibinfo{author}{James~A \surnamestart Momoh\surnameend},
  \bibinfo{author}{ME~\surnamestart El-Hawary\surnameend} \&
  \bibinfo{author}{Ramababu \surnamestart Adapa\surnameend}
  (\bibinfo{year}{1999}): \emph{\bibinfo{title}{A review of selected optimal
  power flow literature to 1993. II. {Newton}, linear programming and interior
  point methods}}.
\newblock {\sl \bibinfo{journal}{IEEE Transactions on Power Systems}}
  \bibinfo{volume}{14}(\bibinfo{number}{1}), pp. \bibinfo{pages}{105--111},
  \doi{10.1109/59.744495}.

\bibitemdeclare{book}{norris1998markov}
\bibitem{norris1998markov}
\bibinfo{author}{James~R \surnamestart Norris\surnameend}
  (\bibinfo{year}{1998}): \emph{\bibinfo{title}{Markov chains}}.
\newblock \bibinfo{volume}{2}, \bibinfo{publisher}{Cambridge university press}.

\bibitemdeclare{article}{rahimi2010demand}
\bibitem{rahimi2010demand}
\bibinfo{author}{Farrokh \surnamestart Rahimi\surnameend} \&
  \bibinfo{author}{Ali \surnamestart Ipakchi\surnameend}
  (\bibinfo{year}{2010}): \emph{\bibinfo{title}{Demand response as a market
  resource under the smart grid paradigm}}.
\newblock {\sl \bibinfo{journal}{IEEE Transactions on Smart Grid}}
  \bibinfo{volume}{1}(\bibinfo{number}{1}), pp. \bibinfo{pages}{82--88},
  \doi{10.1109/tsg.2010.2045906}.

\bibitemdeclare{article}{composableDER2}
\bibitem{composableDER2}
\bibinfo{author}{Lorenzo \surnamestart Reyes-Chamorro\surnameend},
  \bibinfo{author}{Andrey \surnamestart Bernstein\surnameend},
  \bibinfo{author}{Jean-Yves~Le \surnamestart Boudec\surnameend} \&
  \bibinfo{author}{Mario \surnamestart Paolone\surnameend}
  (\bibinfo{year}{2015}): \emph{\bibinfo{title}{A composable method for
  real-time control of active distribution networks with explicit power
  setpoints. Part {II}: Implementation and validation}}.
\newblock {\sl \bibinfo{journal}{Electric Power Systems Research}}
  \bibinfo{volume}{125}, pp. \bibinfo{pages}{265 -- 280},
  \doi{10.1016/j.epsr.2015.03.022}.

\bibitemdeclare{book}{CTincontext}
\bibitem{CTincontext}
\bibinfo{author}{Emily \surnamestart Riehl\surnameend} (\bibinfo{year}{2016}):
  \emph{\bibinfo{title}{Category Theory in Context}}.
\newblock \bibinfo{publisher}{Dover Publications}.
\newblock \urlprefix\url{http://www.math.jhu.edu/~eriehl/context.pdf}.

\bibitemdeclare{article}{peak}
\bibitem{peak}
\bibinfo{author}{Mark~F. \surnamestart Ruth\surnameend},
  \bibinfo{author}{Monte~S. \surnamestart Lunacek\surnameend} \&
  \bibinfo{author}{Birk \surnamestart Jones\surnameend} (\bibinfo{year}{2017}):
  \emph{\bibinfo{title}{Impacts of Using Distributed Energy Resources to Reduce
  Peak Loads in Vermont}}.
\newblock \doi{10.2172/1411137}.

\bibitemdeclare{article}{schultz2016databases}
\bibitem{schultz2016databases}
\bibinfo{author}{Patrick \surnamestart {Schultz}\surnameend},
  \bibinfo{author}{David~I. \surnamestart {Spivak}\surnameend},
  \bibinfo{author}{Christina \surnamestart {Vasilakopoulou}\surnameend} \&
  \bibinfo{author}{Ryan \surnamestart {Wisnesky}\surnameend}
  (\bibinfo{year}{2016}): \emph{\bibinfo{title}{{Algebraic Databases}}}.
\newblock {\sl \bibinfo{journal}{arXiv
  e-prints}}:\bibinfo{eid}{arXiv:1602.03501}.

\bibitemdeclare{article}{sw2017integration}
\bibitem{sw2017integration}
\bibinfo{author}{Patrick \surnamestart Schultz\surnameend} \&
  \bibinfo{author}{Ryan \surnamestart Wisnesky\surnameend}
  (\bibinfo{year}{2017}): \emph{\bibinfo{title}{Algebraic data integration}}.
\newblock {\sl \bibinfo{journal}{Journal of Functional Programming}}
  \bibinfo{volume}{27}, p. \bibinfo{pages}{e24},
  \doi{10.1017/S0956796817000168}.

\bibitemdeclare{inbook}{Selinger2011}
\bibitem{Selinger2011}
\bibinfo{author}{P.~\surnamestart Selinger\surnameend} (\bibinfo{year}{2011}):
  \emph{\bibinfo{title}{A Survey of Graphical Languages for Monoidal
  Categories}}, pp. \bibinfo{pages}{289--355}.
\newblock \bibinfo{publisher}{Springer Berlin Heidelberg},
  \bibinfo{address}{Berlin, Heidelberg}, \doi{10.1007/978-3-642-12821-9_4}.

\bibitemdeclare{article}{spivak2012functorial}
\bibitem{spivak2012functorial}
\bibinfo{author}{David~I. \surnamestart Spivak\surnameend}
  (\bibinfo{year}{2012}): \emph{\bibinfo{title}{Functorial data migration}}.
\newblock {\sl \bibinfo{journal}{Information and Computation}}
  \bibinfo{volume}{217}, pp. \bibinfo{pages}{31 -- 51},
  \doi{10.1016/j.ic.2012.05.001}.

\bibitemdeclare{inproceedings}{pixelarrays}
\bibitem{pixelarrays}
\bibinfo{author}{David~I. \surnamestart Spivak\surnameend},
  \bibinfo{author}{Dominique \surnamestart Ernadote\surnameend} \&
  \bibinfo{author}{Omar \surnamestart Hammammi\surnameend}
  (\bibinfo{year}{2016}): \emph{\bibinfo{title}{Pixel matrices: An elementary
  technique for solving nonlinear systems}}.
\newblock In: {\sl \bibinfo{booktitle}{2016 {IEEE} International Symposium on
  Systems Engineering ({ISSE})}}, \bibinfo{publisher}{{IEEE}},
  \doi{10.1109/syseng.2016.7753120}.

\bibitemdeclare{article}{staymeredith2017opsem}
\bibitem{staymeredith2017opsem}
\bibinfo{author}{Michael \surnamestart Stay\surnameend} \&
  \bibinfo{author}{L.~G. \surnamestart Meredith\surnameend}
  (\bibinfo{year}{2017}): \emph{\bibinfo{title}{Representing operational
  semantics with enriched Lawvere theories}}.
\newblock {\sl \bibinfo{journal}{CoRR}} \bibinfo{volume}{abs/1704.03080}.
\newblock \urlprefix\url{http://arxiv.org/abs/1704.03080}.

\bibitemdeclare{incollection}{Stevens:2007:LBM:1462618.1462630}
\bibitem{Stevens:2007:LBM:1462618.1462630}
\bibinfo{author}{Perdita \surnamestart Stevens\surnameend}
  (\bibinfo{year}{2008}): \emph{\bibinfo{title}{Generative and Transformational
  Techniques in Software Engineering {II}}}.
\newblock chapter \bibinfo{chapter}{A Landscape of Bidirectional Model
  Transformations}, \bibinfo{publisher}{Springer-Verlag},
  \bibinfo{address}{Berlin, Heidelberg}, pp. \bibinfo{pages}{408--424},
  \doi{10.1007/978-3-540-88643-3_10}.

\bibitemdeclare{inproceedings}{trollman2015}
\bibitem{trollman2015}
\bibinfo{author}{Frank \surnamestart Trollmann\surnameend} \&
  \bibinfo{author}{Sahin \surnamestart Albayrak\surnameend}
  (\bibinfo{year}{2015}): \emph{\bibinfo{title}{Extending Model to Model
  Transformation Results from Triple Graph Grammars to Multiple Models}}.
\newblock In \bibinfo{editor}{Dimitris \surnamestart Kolovos\surnameend} \&
  \bibinfo{editor}{Manuel \surnamestart Wimmer\surnameend}, editors: {\sl
  \bibinfo{booktitle}{Theory and Practice of Model Transformations}},
  \bibinfo{publisher}{Springer International Publishing}, pp.
  \bibinfo{pages}{214--229}, \doi{10.1007/978-3-319-21155-8_16}.

\bibitemdeclare{inproceedings}{fire}
\bibitem{fire}
\bibinfo{author}{W.~\surnamestart {Tushar}\surnameend},
  \bibinfo{author}{S.~\surnamestart {Huang}\surnameend},
  \bibinfo{author}{C.~\surnamestart {Yuen}\surnameend}, \bibinfo{author}{J.~A.
  \surnamestart {Zhang}\surnameend} \& \bibinfo{author}{D.~B. \surnamestart
  {Smith}\surnameend} (\bibinfo{year}{2014}): \emph{\bibinfo{title}{Synthetic
  generation of solar states for smart grid: A multiple segment Markov chain
  approach}}.
\newblock In: {\sl \bibinfo{booktitle}{IEEE PES Innovative Smart Grid
  Technologies, Europe}}, pp. \bibinfo{pages}{1--6},
  \doi{10.1109/ISGTEurope.2014.7028832}.

\bibitemdeclare{article}{dmrg}
\bibitem{dmrg}
\bibinfo{author}{Steven~R. \surnamestart White\surnameend}
  (\bibinfo{year}{1992}): \emph{\bibinfo{title}{Density matrix formulation for
  quantum renormalization groups}}.
\newblock {\sl \bibinfo{journal}{Phys. Rev. Lett.}} \bibinfo{volume}{69}, pp.
  \bibinfo{pages}{2863--2866}, \doi{10.1103/PhysRevLett.69.2863}.

\bibitemdeclare{article}{matpower}
\bibitem{matpower}
\bibinfo{author}{Ray~Daniel \surnamestart {Zimmerman}\surnameend},
  \bibinfo{author}{Carlos~Edmundo \surnamestart {Murillo-S\'anchez}\surnameend}
  \& \bibinfo{author}{Robert~John \surnamestart {Thomas}\surnameend}
  (\bibinfo{year}{2011}): \emph{\bibinfo{title}{MATPOWER: Steady-State
  Operations, Planning, and Analysis Tools for Power Systems Research and
  Education}}.
\newblock {\sl \bibinfo{journal}{IEEE Transactions on Power Systems}}
  \bibinfo{volume}{26}(\bibinfo{number}{1}), pp. \bibinfo{pages}{12--19},
  \doi{10.1109/TPWRS.2010.2051168}.

\end{thebibliography}
\end{document}